\documentclass[aps,prb,twocolumn,showpacs,floatfix,superscriptaddress]{revtex4}
\usepackage{dcolumn}
\usepackage{bm}
\usepackage{amsmath,amssymb,graphicx}

\begin{document}
\preprint{Submitted to Phys. Rev. B}

\title{Polarization dependence of coherent phonon generation and detection in highly-aligned single-walled carbon nanotubes}

\author{L.~G.~Booshehri}
\affiliation{Department of Electrical and Computer Engineering, Rice University, Houston, Texas 77005, USA}
\affiliation{The Richard E. Smalley Institute for Nanoscale Science and Technology, Rice University, Houston, Texas 77005}

\author{C.~L.~Pint}
\affiliation{The Richard E. Smalley Institute for Nanoscale Science and Technology, Rice University, Houston, Texas 77005}
\affiliation{Department of Physics and Astronomy, Rice University, Houston, Texas 77005, USA}
\affiliation{Department of Chemistry, Rice University, Houston, Texas 77005, USA}

\author{G.~D.~Sanders}
\affiliation{Department of Physics, University of Florida, Box 118440, Gainesville, Florida 32611-8440}

\author{L.~Ren}
\affiliation{Department of Electrical and Computer Engineering, Rice University, Houston, Texas 77005, USA}
\affiliation{The Richard E. Smalley Institute for Nanoscale Science and Technology, Rice University, Houston, Texas 77005}

\author{C.~Sun}
\affiliation{Department of Electrical and Computer Engineering, Rice University, Houston, Texas 77005, USA}
\affiliation{The Richard E. Smalley Institute for Nanoscale Science and Technology, Rice University, Houston, Texas 77005}

\author{E.~H.~H\'{a}roz}
\affiliation{Department of Electrical and Computer Engineering, Rice University, Houston, Texas 77005, USA}
\affiliation{The Richard E. Smalley Institute for Nanoscale Science and Technology, Rice University, Houston, Texas 77005}

\author{J.-H.~Kim}
\author{K.-J.~Yee}
\affiliation{Department of Physics, Chungnam National University, Daejeon, 305-764, Republic of Korea}

\author{Y.-S.~Lim}
\affiliation{Department of Applied Physics, Konkuk University, Chungju, Chungbuk, 380-701, Republic of Korea}

\author{R.~H.~Hauge}
\affiliation{The Richard E. Smalley Institute for Nanoscale Science and Technology, Rice University, Houston, Texas 77005}
\affiliation{Department of Chemistry, Rice University, Houston, Texas 77005, USA}

\author{C.~J.~Stanton}
\affiliation{Department of Physics, University of Florida, Box 118440,
Gainesville, Florida 32611-8440}

\author{J.~Kono}
\email[]{kono@rice.edu}
\homepage[]{www.ece.rice.edu/~kono}
\thanks{corresponding author.}
\affiliation{Department of Electrical and Computer Engineering, Rice University, Houston, Texas 77005, USA}
\affiliation{The Richard E. Smalley Institute for Nanoscale Science and Technology, Rice University, Houston, Texas 77005}
\affiliation{Department of Physics and Astronomy, Rice University, Houston, Texas 77005, USA}

\date{\today}
\begin{abstract}
We have investigated the polarization dependence of the generation and detection of radial breathing mode (RBM) coherent phonons (CP) in highly-aligned single-walled carbon nanotubes.  Using polarization-dependent pump-probe differential-transmission spectroscopy, we measured RBM CPs as a function of angle for two different geometries. In Type I geometry, the pump and probe polarizations were fixed, and the sample orientation was rotated, whereas, in Type II geometry, the probe polarization and sample orientation were fixed, and the pump polarization was rotated. In both geometries, we observed a very nearly complete quenching of the RBM CPs when the pump polarization was perpendicular to the nanotubes.  For both Type I and II geometries, we have developed a microscopic theoretical model to simulate CP generation and detection as a function of polarization angle and found that the CP signal decreases as the angle goes from 0$^\circ$ (parallel to the tube) to 90$^\circ$ (perpendicular to the tube).  We compare theory with experiment in detail for RBM CPs created by pumping at the $E_{44}$ optical transition in an ensemble of single-walled carbon nanotubes with a diameter distribution centered around 3~nm, taking into account realistic band structure and imperfect nanotube alignment in the sample.
\end{abstract}

\pacs{78.67.Ch, 63.22.+m, 73.22.-f, 78.67.-n}
\maketitle

\section{Introduction}

The one-dimensionality of single-walled carbon nanotubes (SWNTs) is attractive from both fundamental and applied points of view, where the 1D confinement of electrons and phonons results in unique anisotropic electric, magnetic, mechanical, and optical properties.\cite{Dresselhaus:2003book,Charlier07.643} Individualized SWNTs, both single-tube and in ensemble samples, have shown anisotropy with polarized Raman scattering and optical absorption measurements where maximum signals result when the nanotube axis is aligned parallel to the polarization of incident light.\cite{Gommans:2000p17979,Duesberg:2000p18195, Jorio:2000p18204, Li:2001p18206, Hartschuh:2003p18205, Islam:2004p18212} Additionally, due to strong anisotropic magnetic susceptibilities, both semiconducting and metallic SWNTs align well within an external magnetic field, and with the added properties of the Aharonov-Bohm effect, the electronic band structure of SWNTs respond anisotropically with the strength of a tube-threading magnetic flux.\cite{Ajiki:1993p18207, Lu:1995p18208, Marques:2006p18209, Zaric:2004p17316, Zaric:2004p17447, Islam:2005p18214, Shaver:2009p18210, Searles:2010p18190} Such optical and magnetic anisotropy is also expected with bulk samples, but detailed measurements showing extreme anisotropy have been lacking.\cite{Ren:2009p18188}

Here we investigate anisotropic optical and vibrational properties of a bulk film of highly-aligned SWNTs using polarization-dependent coherent phonon (CP) spectroscopy.  CP spectroscopy is an ultrafast pump-probe technique that complements CW Raman spectroscopy, and although both techniques provide information about electron-phonon coupling, CP spectroscopy avoids the common disadvantages of Raman spectroscopy that include detection of Rayleigh scattering and photoluminescence and the broadening and blending of peak features.\cite{Lim:2006p12950} Such advantages are useful when investigating SWNTs, where a large majority of samples include ensembles dispersed in various environments and their optical properties are obscured by the collection of varying species of nanotubes.

Recent CP studies on SWNTs have produced direct observation of CP oscillations of both the radial breathing mode (RBM) and G-band phonons, in addition to their phase information and dephasing times.\cite{Lim:2006p12950,Gambetta:2006p3342,Kato:2008p12903,Yee:2009p16763,Lim:2009p16762,Kim:2009p15417,LimetAl10ACS} Furthermore, when pulse-shaping techniques are combined with CP spectroscopy, predesigned trains of femtosecond pulses selectively excite RBM CPs of a specific chirality, avoiding inhomogeneous broadening from the ensemble.\cite{Kim:2009p15417} However, it is important to note that previous CP studies investigated randomly-aligned SWNT samples, and as the quasi-1D nature of SWNTs leads to optical anisotropy that is dominant in the polarization dependence of PL, absorption, and Raman scattering, CP studies on aligned SWNTs are necessary.

Below, we present results of a detailed experimental and theoretical study of CPs in highly-aligned SWNT thin films and found a very strong polarization anisotropy of the RBM as a function of angle. In particular, we observed a very nearly complete quenching of the RBM when the optical polarization is perpendicular to the tubes.  We have developed a theoretical model to understand this extreme anisotropy, including band structure, interband optical transition elements, ultrafast carrier and phonon dynamics, and imperfect nanotube alignment in the sample.  Fitting our results to theory, we also calculated the nematic order parameter, $S$,\cite{Shaver:2009p18210,Searles:2010p18190,Ren:2009p18188} i.e., the degree of alignment of SWNTs in the sample.

\section{Samples and Experimental Methods}

\begin{figure}
\includegraphics[scale=0.75]{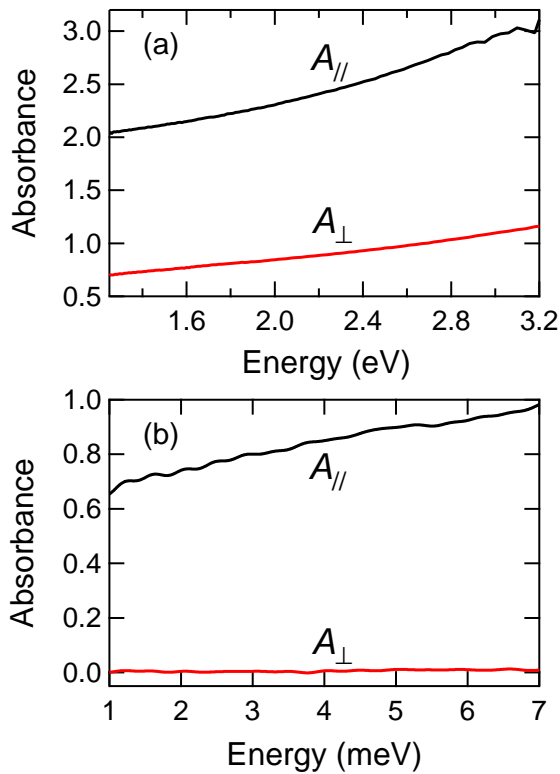}
\caption{\label{fig1}(color online) Absorption spectra of aligned single-walled carbon nanotubes in the (a) near-infrared to visible and (b) far-infrared (or terahertz) ranges for parallel ($A_{\parallel}$) and perpendicular ($A_{\perp}$) polarization.}
\end{figure}
\begin{figure} [tbp]
\includegraphics[width=2.5in]{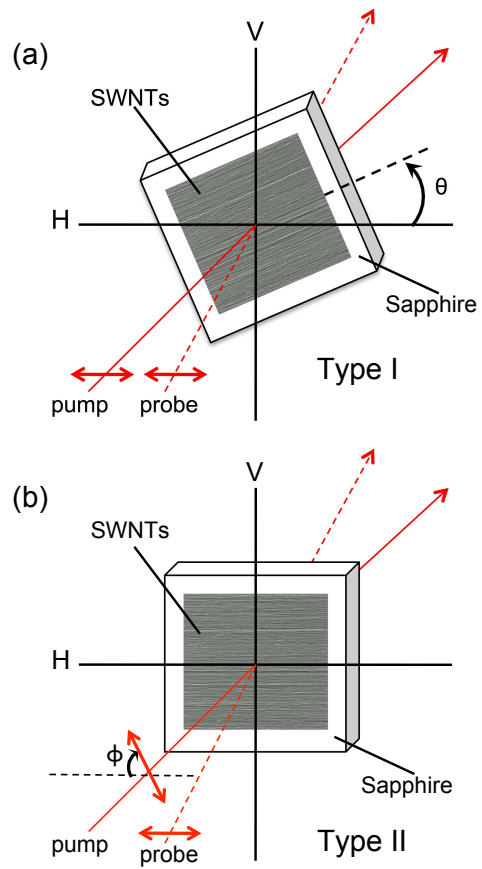}
\caption{\label{fig2}(color online) Scanning electron microscopy image of aligned single-walled carbon nanotubes, and the two experimental configurations employed in the current pump-probe spectroscopy work.  (a) Type I: pump and probe polarizations are fixed and sample orientation is rotated. (b) Type II: probe and sample orientations are fixed and pump polarization is rotated.}
\end{figure}

We investigated CPs through degenerate pump-probe spectroscopy measurements on highly-aligned SWNT thin films transferred onto sapphire substrates.  Patterned, vertically aligned SWNT arrays grown by chemical vapor deposition were subsequently etched with H$_2$O vapor to allow transffer to sapphire via a dry contact transfer printing technique, which forms horizontally-aligned SWNT thin films.\cite{Pint:2008p17957,Pint:2010p18012} The resulting film produces aligned nanotubes of the same length, with a diameter distribution centered around 3~nm.\cite{Pint:2010p18012} Figure~\ref{fig1}(a) shows polarization-dependent absorption spectra of a typical aligned sample in the visible to near-infrared range, while Fig.~\ref{fig1}(b) shows absorption spectra in the far-infrared or terahertz (THz) range. The polarization anisotropy is obvious in both (a) and (b), but extreme in the THz regime (b), as there is virtually zero absorption when the sample is perpendicular to the THz polarization.\cite{Ren:2009p18188}  Therefore, with such an aligned sample, our polarization measurements can be extended to include the sample as an additional rotation parameter, compared to our earlier study of CP polarization dependence in randomly-aligned SWNTs.\cite{Yee:2009p16763}

Using a mode-locked Ti:Sapphire laser with $\sim$80~fs pulse width that is shorter than the period of excited CP oscillations and $\sim$40~mW average pump power, the laser was tuned to central wavelength of 850~nm (1.46~eV) to predominantly excite the $E_{44}$ interband transitions of the SWNTs in the sample.  A shaker-delay system and balance detector was employed for fast-scan, real-time observation of CPs. Extraction of CP oscillations from raw pump-probe time-domain signal were performed as previously described.\cite{Lim:2006p12950} As depicted in Fig.~\ref{fig2}, two types of polarization measurements were investigated. Type I configuration maintained the same polarization for the pump and probe, while the alignment axis of the sample was rotated. Type II configuration maintained the same orientation for the probe and sample, while the pump polarization was rotated.  For Type II measurements, a half-wave plate provided 90-degree rotation of the pump.  All measurements were performed at room temperature.

\section{Experimental Results}

\begin{figure}
\includegraphics[scale=0.54]{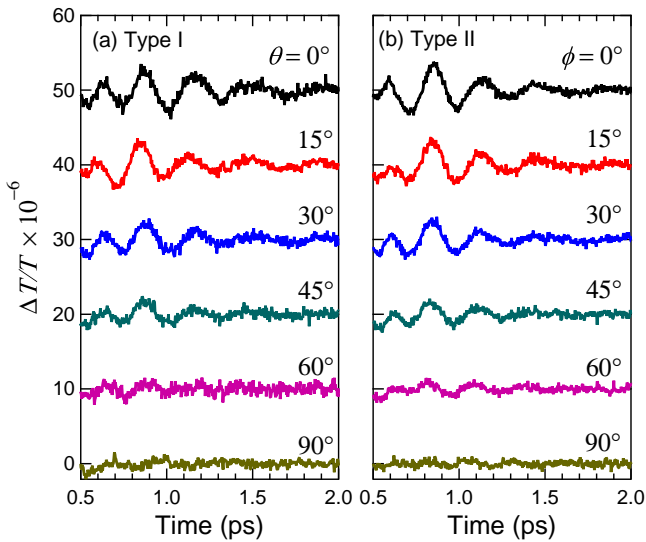}
\caption{\label{fig3}(color online) Experimental differential transmission data in the time domain showing coherent phonon oscillations of the radial breathing mode for different polarization angles in (a) Type I and (b) Type II configurations (see Fig.~\ref{fig1}).  The traces are vertically offset for clarity.}
\end{figure}

Results of our polarization-dependent ultrafast pump-probe differential transmission measurements on the aligned SWNT film for both Type I and Type II orientations are shown in Fig.~\ref{fig3}. Here, it is seen that the typical differential transmission amplitude of the RBM CP oscillations is on the order of 10$^{-6}$, with a CP decay time of $\sim$1.5~ps.  This short decay time, compared to our earlier studies on individually-suspended SWNTs in solution,\cite{Lim:2006p12950,Yee:2009p16763,Kim:2009p15417} is expected with our sample of bundled SWNTs.  As the polarization angle is rotated, we see a strong polarization anisotropy of the RBM CPs as a function of angle, where the strongest oscillations are observed at 0$^\circ$ while the signal is very nearly completely quenched at 90$^\circ$ for both Type I and Type II geometries.

\begin{figure}
\includegraphics[scale=0.75]{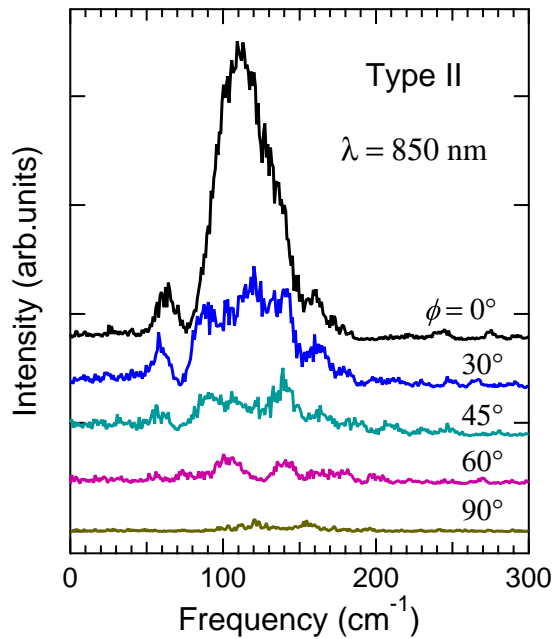}
\caption{\label{fig4}(color online) Coherent phonon spectra for different polarization angles in Type II configuration obtained through Fourier transform of the time-domain data in Fig.~\ref{fig3}(b). The traces are vertically offset for clarity.}
\end{figure}

\begin{figure}
\includegraphics[scale=0.62]{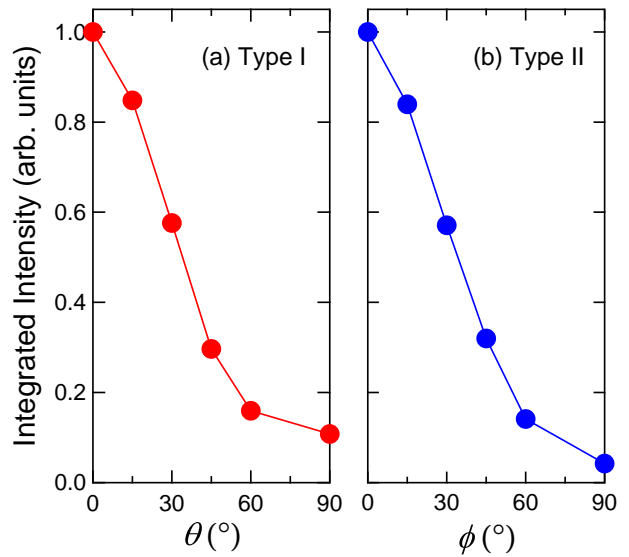}
\caption{\label{fig5}(color online) Spectrally integrated coherent phonon intensity as a function of angle, measured in (a) Type I and (b) Type II configurations.}
\end{figure}

Figure~\ref{fig4} shows CP spectra for different polarization angles for Type II configuration.  To produce these spectra in the frequency domain, we calculated the Fourier transform (FT) of the time-domain CP oscillations.  The frequencies of the CP oscillations have a wide range, from 50 to 200~cm$^{-1}$, which is consistent with the large diameter distribution of nanotubes within the sample.\cite{Pint:2010p18012} The polarization anisotropy is clearly observed in the CP spectra.  Figures~\ref{fig5}(a) and \ref{fig5}(b) plot the integrated CP intensity of the FT for all excited nanotubes as a function of $\theta$ [in Fig.~\ref{fig5}(a)] and $\phi$ [in Fig.~\ref{fig5}(b)], where $\theta$ ($\phi$) is the angle of rotation for the Type I (Type II) configuration.


\section{Theory}

We have developed a microscopic theory for the generation of coherent phonons in single-walled carbon nanotubes and their detection in coherent phonon spectroscopy experiments.  Our microscopic theory is described in detail in Ref.~\onlinecite{Lim:2009p16762}, so we only summarize the main points here and indicate how our earlier work has been extended to include polarization-dependent coherent phonon spectroscopy.

We treat the $\pi$ and $\pi^*$ electronic states in ($n$,$m$) carbon nanotubes using a third-nearest-neighbor extended tight binding (ETB) formalism developed by Porezag \textit{et al}.\cite{Porezag:1995p18010} for carbon compounds. Using a density functional based parametrization, Porezag \textit{et al.} derived analytic expressions for the Hamiltonian and overlap matrix elements that depend only on the C-C bond lengths.  Using the ETB formalism, we obtain tight-binding wave functions and electronic energy levels $E_{s \mu}(k)$ where $s = c,v$ labels the conduction and valence bands, $\mu$ labels the cutting lines, and $k$ is the one-dimensional nanotube Brillouin zone.\cite{Dresselhaus:2003book}  By exploiting the screw symmetry of the nanotube, we can block-diagonalize the electronic Hamiltonian and overlap matrices into $2 \times 2$ blocks, one for each cutting line, as described in Appendix A of Ref.~\onlinecite{Lim:2009p16762}.

We treat nanotube lattice dynamics using a seven-parameter valence-force-field model described in Appendix B of Ref.~\onlinecite{Lim:2009p16762}.  Exploiting the nanotube screw symmetry, the dynamical matrix can be block-diagonalized into $6 \times 6$ blocks, one for each cutting line, which can be solved for the phonon displacement vectors and dispersion relations $\omega^2_{\beta\nu}(q)$.  In the the dispersion relations, $\beta = 1, \cdots, 6$ labels the phonon modes, $\nu$ is an angular momentum quantum number, and $q$ is the phonon wave vector in the one-dimensional nanotube Brillouin zone.  Following the work of Jiang \textit{et al}.\cite{Jiang:2006p18013} and Lobo and Martins,\cite{Lobo:1997p159} we include four types of force field potentials, i.e., the bond stretching, in-plane bond bending, out-of-plane bond bending, and bond twisting potentials.  Our force-field potential energies are invariant under rigid rotations and translations (force constant sum rule), and, as a result, our phonon model correctly predicts the dispersion relation for the long-wavelength flexure modes.\cite{Lim:2009p16762}

The electron-phonon interaction is treated in a second-quantized formalism where the electron-phonon interaction matrix elements are evaluated in Appendix C of Ref.~\onlinecite{Lim:2009p16762}.  In calculating electron-phonon matrix elements, we use 2$p_z$ graphene atomic wave functions and screened atomic potentials obtained from an \textit{ab initio} calculation.\cite{Jiang:2006p18013}

We obtain equations of motion for coherent phonon amplitudes using the Heisenberg equations as described by Kuznetsov and Stanton.\cite{Kuznetsov:1994p18011}  We assume that the optical pulse and photoexcited carrier distributions are distributed uniformly over the nanotube.  In this case, only the $\nu = q = 0$ phonon modes are excited.  To excite RBM coherent phonons, the laser pulse must be short in comparison with the RBM phonon oscillation period.  For coherent RBM phonons, the coherent phonon amplitude is proportional to the tube diameter $D(t)$.  The tube diameter, being proportional to the coherent phonon amplitude, satisfies a driven oscillator equation
\begin{equation}
\frac{\partial^2 D(t)}{\partial t^2} + \omega^2 D(t)= S(t)
\label{RBM Diameter Equation}
\end{equation}
where $\omega$ is the angular frequency of the $\nu = q = 0$ RBM phonon mode.  The driving function $S(t)$ for RBM coherent phonons is given by
\begin{equation}
S(t) \propto - \sum_{s \mu k} M_{s \mu}(k) \left[ f_{s \mu}(k,t) - f^0_{s \mu}(k) \right]
\label{S(t) Equation}
\end{equation}
where $M_{s \mu}(k)$ is the driving function kernal for RBM coherent phonons defined in Eq.~(12) of Ref.~\onlinecite{Lim:2009p16762} and $f_{s \mu}(k,t)$ and $f^0_{s \mu}(k)$ are the time-dependent and initial equilibrium carrier distribution functions, respectively.

From Eq.~(\ref{S(t) Equation}) we see that the driving function depends on the photoexcited carrier distribution functions.  We treat photoexcitation of carriers in a Boltzmann equation formalism and obtain the photogeneration rate using Fermi's golden rule.  In Ref.~\onlinecite{Lim:2009p16762} we only considered linearly-polarized laser pulses with the electric polarization vector parallel to the nanotube axis.  For parallel polarization, optical transitions
only occur between states with the same cutting line index $\mu$, and the resulting photogeneration rate is given by Eq.~(13) in Ref.~\onlinecite{Lim:2009p16762}.

In the present work, we consider linearly-polarized pump and probe beams in which the electric polarization vector makes a finite angle with the nanotube axis.  If $\hat{\epsilon}$ is the unit electric polarization vector for the pump, then the photogeneration rate is now given by
\begin{eqnarray}
\nonumber &&
\left. \frac{\partial f_{s \mu}(k)}{\partial t} \right|_{\mbox{gen}} =
\frac{8 \pi^2 e^2 \ u(t)}{\hbar \ n_g^2 \ (\hbar\omega)^2}
\left(\frac{\hbar^2}{m_0} \right) \sum_{s' \mu'}
\arrowvert \hat{\epsilon} \cdot \vec{P}^{\mu \mu'}_{s s'}(k) \arrowvert^2
\\ &&
\times \Big( f_{s' \mu'}(k,t) - f_{s \mu}(k,t) \Big)
\ \delta \Big( \Delta E^{\mu \mu'}_{s s'}(k) - \hbar\omega \Big)
\label{photogeneration rate}
\end{eqnarray}
where $\Delta E^{\mu \mu'}_{s s'}(k) = \arrowvert E_{s \mu}(k) - E_{s' \mu'}(k) \arrowvert$ are the $k$-dependent transition energies, $\hbar\omega$ is the pump photon energy, $u(t)$ is the time-dependent energy density of the pump pulse, $e$ is the electronic charge, $m_0$ is the free electron mass, and $n_g$ is the index of refraction in the surrounding medium.  To account for spectral broadening of the laser pulses, the delta function in Eq.~(\ref{photogeneration rate}) was replaced with a Lorentzian lineshape\cite{Chuang:1995book}
\begin{equation}
\delta(\Delta E-\hbar\omega) \rightarrow
\frac{\Gamma_p /(2\pi)} {{(\Delta E-\hbar\omega)^2+(\Gamma_p/2)^2}}
\label{delta function broadening}
\end{equation}

In Eq.~(\ref{photogeneration rate}), $\vec{P}^{\mu \mu'}_{s s'}(k)$ are the dipole-allowed optical matrix elements between the initial and final states where $\mu$ and $\mu'$ are initial and final state cutting lines.  For polarization parallel to the tube axis (taken to be $\hat{z}$), $\mu'=\mu$ and $\hat{z} \cdot \vec{P}^{\mu \mu}_{s s'}(k) = P^{\mu}_{s s'}(k)$ with the right hand side being defined in Eq.~(14) of Ref.~\onlinecite{Lim:2009p16762}. For linear polarization parallel to the $x$ axis, $\mu'=\mu \pm 1$ and in the notation of Ref.~\onlinecite{Lim:2009p16762}, we have
\begin{eqnarray}
\nonumber &&
\hat{x} \cdot \vec{P}^{\mu,\mu \pm 1}_{s s'}(k) = \frac{\hbar}{\sqrt{2 m_0}}
\ \frac{1}{2} \sum_{r'} C^{*}_{r'}(s',\mu \pm 1, k)
\\ \nonumber &&
\times \sum_{r\textbf{J}}
\ C^{}_r(s, \mu, k) \ e^{i \phi_\textbf{J}(k,\mu)} \
\big( M_x(r',r\textbf{J}) \pm i M_y(r',r\textbf{J}) \big)
\\
\label{Px equation}
\end{eqnarray}
Similarly, for linear polarization along the $y$ axis, we have
\begin{equation}
\hat{y} \cdot \vec{P}^{\mu,\mu \pm 1}_{s s'}(k) =
\mp \ i \big( \hat{x} \cdot \vec{P}^{\mu,\mu \pm 1}_{s s'}(k) \big)
\label{Py equation}
\end{equation}
In Eqs.~(\ref{Px equation}) and (\ref{Py equation}), the atomic dipole matrix element vectors (which can be evaluated analytically) are given by
\begin{equation}
\bf{M}(r',r\textbf{J}) = \int d\textbf{r} \
\varphi^{*}_{r'\textbf{0}}(\textbf{r}-\textbf{R}_{r'\textbf{0}}) \
\nabla \
\varphi_{r\textbf{J}}(\textbf{r}-\textbf{R}_{r\textbf{J}})
\label{Mxyz equation}
\end{equation}
where the $2p_z$ orbitals $\varphi_{r\textbf{J}}$ are defined in Eq.~(C3) of Ref.~\onlinecite{Lim:2009p16762}.

In our experiments, a probe pulse is used to measure the time-varying absorption coefficient.  In our theory, the
time-varying optical properties measured by the probe pulse are obtained from the imaginary part of the dielectric function. For a linearly polarized probe pulse, we have
\begin{eqnarray}
\label{imaginary dielectric function}
\nonumber &&
\varepsilon_2(\hbar\omega,t) =
\frac{8 \pi^2 e^2}{A_t (\hbar\omega)^2}
\left( \frac{\hbar^2}{m_0} \right) \sum_{s s' \mu \mu' }
\int \frac{dk}{\pi}
\ \arrowvert \hat{\epsilon} \cdot \vec{P}^{\mu \mu'}_{s s'}(k) \arrowvert^2
\\ &&
\times \Big( f_{s \mu}(k,t) - f_{s' \mu'}(k,t) \Big)
\ \delta \Big( \Delta E^{\mu \mu'}_{s s'}(k) - \hbar\omega \Big)
\end{eqnarray}
where $A_t=\pi (d_t/2)^2$ is the cross-sectional area of the tube and $d_t$ is the equilibrium nanotube diameter.  The dirac delta function is replaced by a broadened Lorentzian of the form shown in Eq.~(\ref{delta function broadening}). The photon energy of the probe pulse photons is $\hbar\omega$, and $\hat{\epsilon}$ is the probe unit polarization vector. In our experiments, the photon energy $\hbar\omega$ is the same for both pump and probe, while the polarization vector $\hat{\epsilon}$ may be different for pump and probe.  To obtain the computed coherent phonon spectrum, we take the power spectrum of the computed differential transmission signal after background subtraction using the Lomb periodogram algorithm described in Ref.~\onlinecite{Press:1992book}.

\section{Comparison of theory and experiment}


To compare experiment with theory, we performed simulations of polarization-dependent CP spectroscopy on a (38,0) zigzag nanotube.  This is a mod-2 semiconducting tube with a diameter of 3.01~nm.  Our sample contains an ensemble of nanotubes with diameters centered around 3~nm,\cite{Pint:2010p18012} and we expect that (38,0) tubes will contribute
strongly to the CP signal for the ensemble since (i) the (38,0) tube has a diameter in the center of the measured diameter distribution and (ii) mod-2 zigzag tubes tend to have very strong intrinsic CP signals.\cite{Lim:2006p12950,Kim:2009p15417,Lim:2009p16762}
%
\begin{figure} [tbp]
\includegraphics[scale=1.0]{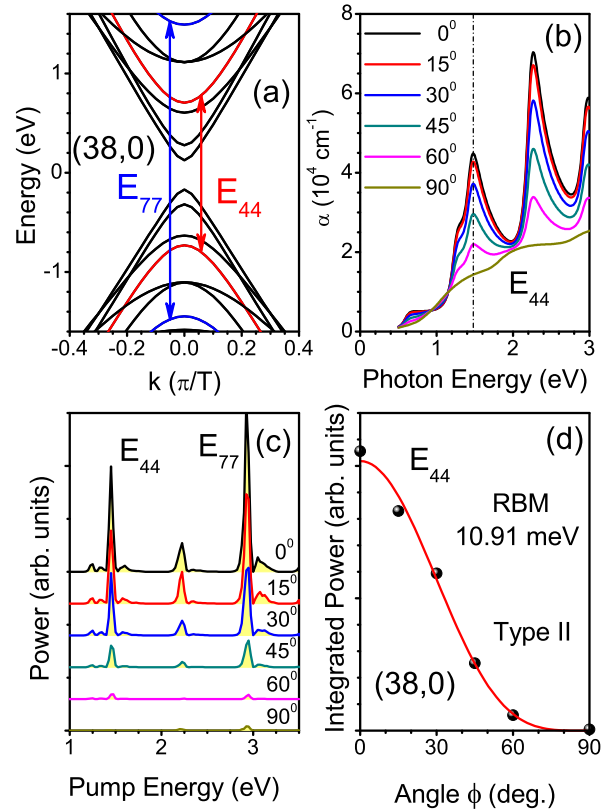}
\caption{\label{Theoretical Type II Results 4 Panel}(color online) (a) Theoretical bandstructure for (38,0) nanotubes showing the strong $E_{44}$ and $E_{77}$ transitions for a polarization vector parallel to the tube. The wavevector $k$ is expressed in units of $\pi / T$ where $T = 4.31 \ \AA$ is the length of the translational unit cell. (b) Computed absorption spectra for polarization angles varying from 0$^\circ$ (parallel to tube) to 90$^\circ$ (perpendicular to tube).
(c) Coherent phonon spectra as a function of pump energy and pump polarization angle $\phi$
for RBM coherent phonons ($\omega = 10.91 \ \mbox{meV}$). Coherent phonon spectra
are Fourier transforms of computed time dependent differential transmission for Type II
pump-probe experiments. (d) Integrated power (black dots) obtained by taking the
area under the computed $E_{44}$ peaks in lower left panel. We fit (red curve) the
theoretical calculations to $A \cos^p(\phi)$ where $A$ and $p = 4.017$ are
fitting parameters.}
\end{figure}
%

Our theoretical results for the (38,0) nanotube are shown in Fig.~\ref{Theoretical Type II Results 4 Panel}.  Our computed electronic $\pi$ band structure for the (38,0) tube is shown in Fig.~\ref{Theoretical Type II Results 4 Panel}(a).  The bands are doubly degenerate with two distinct cutting lines corresponding to each band.  The $\pi$ valence bands have negative energy, and the $\pi^*$ conduction bands have positive energy.  For $z$-polarized light, the allowed $E_{44}$ and $E_{77}$ transitions (selection rule $\Delta\mu = 0$) giving rise to the strongest CP signals are indicated by vertical arrows.

The absorption coefficient for (38,0) nanotubes as a function of linearly-polarized photon energy is shown in Fig.~\ref{Theoretical Type II Results 4 Panel}(b) for polarization angles varying from 0$^\circ$ (parallel to tube) to 90$^\circ$ (perpendicular to tube).

For the (38,0) nanotube, the RBM phonons have a computed energy $\hbar\omega = 10.91 \ \mbox{meV}$ and an oscillation period of 379~fs.  We simulate the generation of RBM coherent phonons by pumping with a 50~fs pump pulse, which is much shorter than the RBM oscillation period.  Polarization-dependent CP spectra for RBM coherent phonons are shown in Fig.~\ref{Theoretical Type II Results 4 Panel}(c) for pump polarization angles $\phi$ ranging from 0$^\circ$ to 90$^\circ$.  In these simulations, the probe polarization is kept fixed parallel to the tube axis (Type II geometry).  As can be seen in the figure, the CP signal is maximum when the pump is polarized parallel to the tube.  As the pump photon energy is varied, we excite RBM coherent phonons by successively photoexciting carriers in different bands.  The strongest RBM CP signals are obtained by pumping at the $E_{44}$ and $E_{77}$ transitions.  As the pump polarization angle increases, the RBM CP signal decreases until it is finally quenched when the pump polarization is perpendicular to the tube.

By taking the area under the RBM CP signal curves in the vicinity of the computed $E_{44}$ transition energy near 1.48~eV, we obtain the integrated power as a function of pump polarization angle shown as black dots in Fig.~\ref{Theoretical Type II Results 4 Panel}(d).  The integrated CP power can be well fit with a fitting function of the form $A\cos^p(\theta)$, where $A$ and $p = 4.017$ are fitting parameters.
%
\begin{figure} [tbp]
\includegraphics[scale=0.75]{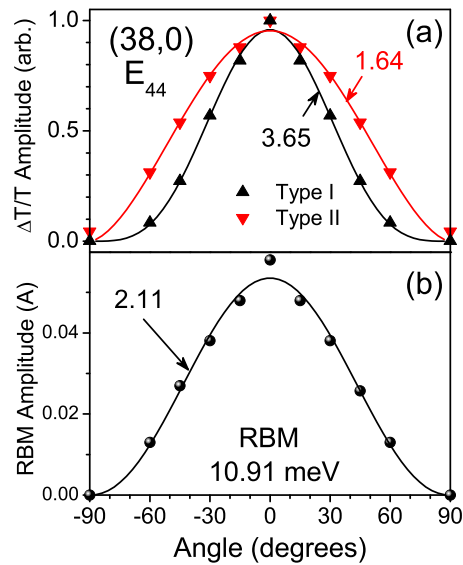}
\caption{\label{DT and RBM vs Angle}(color online)
In (b) we plot the amplitude of RBM coherent phonon oscillations as a function of the pump polarization angle ($\theta$ or $\phi$) for a (38,0) nanotube photoexcited by a $50 \ \mbox{fs}$ pump at the theoretical $E_{44}$ transition. In (a) we plot the amplitude of the resulting differential transmission signal for Type I and II experiments as a function of angles $\theta$ and $\phi$ respectively. Our results are fit to $A \cos^p(\theta)$ $[A \cos^p(\phi)]$ where numerical values of the best fit $p$ are indicated in the figure.}
\end{figure}
%

The polarization dependence of the RBM CP integrated power in Fig.~\ref{Theoretical Type II Results 4 Panel}(d) can be understood by examining Fig.~\ref{DT and RBM vs Angle}. In Fig.~\ref{DT and RBM vs Angle}(b) we plot the coherent RBM diameter oscillations as a function of the pump polarization angle $\theta$. Fitting the theoretical results to $A \cos^p(\theta)$, we obtain $p = 2.11$ as indicated in the figure. For the RBM diameter oscillations we get $p \sim 2$ which makes sense. The driving function $S(t)$ is proportional to the photogenerated carrier density as seen in Eq.~(\ref{S(t) Equation}) and this in turn is proportional to the squared optical matrix element which has a $\cos^2(\theta)$ angular dependence\cite{Gommans:2000p17979}. In Fig.~\ref{DT and RBM vs Angle}(a) we plot the amplitude of the time-dependent differential transmission oscillations measured by the probe beam after background subtraction. For the differential transmission signal, we get $p \sim 2$ for Type II and $p \sim 4$ for Type I. This also makes sense. In Type II experiments the probe polarization is parallel to the tube axis and the pump polarization angle $\theta$ is varied. In this case the signal should be proportional to the amplitude of the diameter oscillations. In Type I experiments on the other hand, the pump and probe polarizations are parallel and we should get an extra factor of $\cos^2(\phi)$ to account for anisotropy of the probe beam absorption. In our CP spectroscopy measurements, we extract the power spectrum of the differential transmission oscillations. The CP power spectrum should be proportional to the square of the amplitude of the differential transmission signal which we anticipate will give us a $\cos^4(\phi)$ polarization dependence for Type II CP spectroscopy experiments and a $\cos^8(\theta)$ dependence for Type I CP spectroscopy experiments.
%
\begin{figure} [tbp]
\includegraphics[scale=0.75]{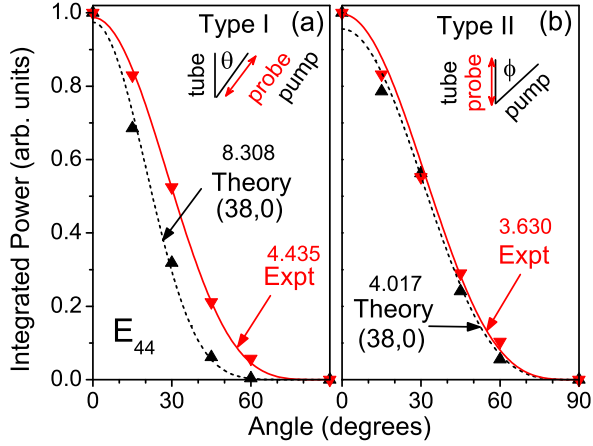}
\caption{\label{CP Experiment Theory Comparison}(color online)
Integrated RBM coherent phonon power as a function of (a) $\theta$ for Type I and (b) $\phi$ for Type II experiments pumping close to the $E_{44}$ transition. Experimental data points are red downward pointing triangles fit by a solid red line. Theoretical data points for a (38,0) nanotube are black upward pointing triangles and are fit by a black dashed line. The fitting functions are of the form $A \cos^p(\theta)$ and $A \cos^p(\phi)$ where the numerical values of the best fit $p$ are indicated in the figure. The experimental curves have undergone background subtraction to obtain complete quenching at an angle of $90^{\circ}$ and then rescaled so the
integrated power at an angle of $0^{\circ}$ is equal to unity.}
\end{figure}
%

The experimental integrated CP power is obtained by taking the Fourier transform power spectrum of the time-dependent differential transmission data shown in Fig.~\ref{fig3} and then computing the area under the $E_{44}$ peak in the resulting spectrum.  The results for our Type I and Type II experiments are shown in Fig.~\ref{CP Experiment Theory Comparison}, where the experimental data points are plotted as downward-pointing red triangles and the corresponding theoretical predictions are shown as upward-pointing black triangles.  In overlaying the experimental and theoretical data points, we subtracted a background from the experimental data to obtain complete quenching at 90$^{\circ}$.  We then rescaled experimental and theoretical data (both in arbitrary units) so that the integrated power at 0$^{\circ}$ is equal to unity.  We then fit experimental and theoretical data to functions of the form $A\cos^p(\theta)$ [$A\cos^p(\phi)$]. The best fit functions for our experimental and theoretical results are shown in Fig.~\ref{CP Experiment Theory Comparison} as solid red and dashed black lines, respectively.

For the Type II experiments, where the probe polarization is fixed parallel to the average tube orientation, we get decent agreement between theory and experiment with best fit parameters $p = 4.017$ for theory and $p = 3.630$ for experiment.  For the Type I experiments, where pump and probe polarization vectors are parallel to each other, there is a discrepancy between theory and experiment.  As can be seen in Fig.~\ref{CP Experiment Theory Comparison}(a), the best fit parameters are $p = 8.308$ for theory and $p = 4.435$ for experiment.  The experimental fits in Fig.~\ref{CP Experiment Theory Comparison} would imply that the CP intensity scales roughly as $\cos^4(\theta)$ independently of the probe polarization.  This seems unlikely given the anisotropy of the absorption coefficient in carbon nanotubes.  In fact, from our theory, we would expect $p \sim 4$ for Type II and $p \sim 8$ for Type I.

We believe that the cause of this discrepancy is most likely due to misalignment effects.  Our sample consists of an ensemble of nanotubes lying horizontally on a sapphire substrate.  While the tubes are highly-aligned, they are not perfectly aligned.  To consider the effects of misalignment on the experimentally-measured integrated CP intensity in a Type I experiment, we assume that the tube alignment angles on the transferred film are described by a Gaussian distribution function with a small standard deviation $\Delta\theta$.  If the angle between the pump polarization vector and the ensemble averaged tube axis is $\theta$, then the angles $\vartheta$ between the pump polarization vector and the axes of each tube in the ensemble are described by a Gaussian distribution
\begin{equation}
P(\vartheta,\theta,\Delta\theta)= \frac{1}{\sqrt{2 \pi} (\Delta\theta)}
\exp{ \left( -\frac{ (\vartheta-\theta)^2 }{2(\Delta\theta)^2} \right) }
\label{Gaussian Distribution Function}
\end{equation}

If the integrated RBM coherent phonon power for each nanotube in the ensemble is given by $A \cos^p(\vartheta)$, the ensemble averaged integrated power $I_{cp}(\theta,\Delta\theta)$ is obtained by taking the ensemble average over $\vartheta$
\begin{equation}
I_{cp}(\theta,\Delta\theta) = A \ \int_{-\infty}^{\infty} d\vartheta \
P(\vartheta,\theta,\Delta\theta) \ \cos^p(\vartheta)
\label{Averaged CP Power}
\end{equation}
In Eq.~(\ref{Averaged CP Power}), we extended the integration limits on $\vartheta$ to infinity in the limit of small $\Delta\theta$.  If $p$ is a positive integer, $I_{cp}(\theta,\Delta\theta)$ can be found analytically.  Otherwise, it can be evaluated numerically. In the case of Type II experiments, the ensemble averaged integrated power is obtained by replacing $\theta$ with $\phi$ in Eqs.~(\ref{Gaussian Distribution Function}) and (\ref{Averaged CP Power}).
%
\begin{figure} [tbp]
\includegraphics[scale=0.75]{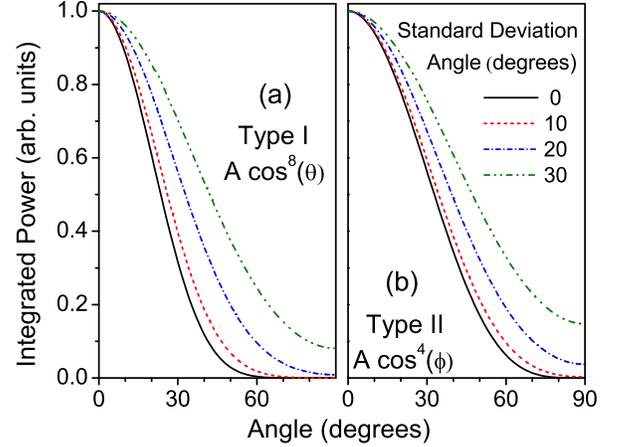}
\caption{\label{Misalignment Effects}(color online) Effects of tube misalignment on integrated coherent phonon power for an ensemble of nanotubes with tube axis orientation angles following a Gaussian distribution. (a) Type I fitting function $A \cos^p(\theta)$ with $p = 8$ and (b) Type II fitting function $A \cos^p(\phi)$ with $p = 4$.}
\end{figure}

\begin{figure} [tbp]
\includegraphics[scale=0.75]{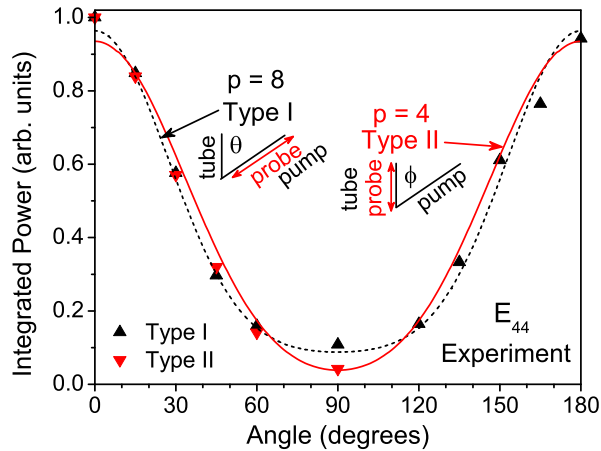}
\caption{\label{Experimental Fit}(color online) Experimental integrated RBM coherent phonon power for the $E_{44}$ transition as a function of $\theta$ for Type I
(black upward pointing triangles) and $\phi$ for Type II (red downward pointing triangles) experiments. Type I experiments are fit to $A \cos^p(\theta + \Delta \theta) + B$ where $p = 8$ (black dashed line) and Type II experiments are fit to $A \cos^p(\phi + \Delta \phi) + B$ for $p = 4$ (solid red line). $A$ and $B$ are background subtraction and rescaling parameters and the standard deviations for the random tube
axis misalignment $\Delta \theta$ and $\Delta \phi$ are restricted to be the same for both Type I and II.}
\end{figure}

The effects of nanotube misalignment on the integrated CP power fitting function for an ensemble of Gaussian misaligned tubes is illustrated in Fig.~\ref{Misalignment Effects}, assuming $p = 8$ for Type I and $p = 4$ for Type II experiments.  As the standard deviations $\Delta \theta$ [$\Delta \phi$] increase, quenching of the CP intensity is reduced and $I_{cp}(\theta,\Delta\theta)$ [$I_{cp}(\phi,\Delta\phi)$] become flatter.
%

Using this model, we are able to get reasonable fits to the experimentally measured CP intensity in both Type I and Type II experiments.  We assume a fitting function of the form $A\cos^p(\theta+\Delta\theta) + B$ [$A\cos^p(\phi+\Delta\phi) + B$], where $A$ and $B$
are background subtraction and rescaling parameters and $\Delta \theta$ [$\Delta \phi$] is a random Gaussian distributed misalignment angle (standard deviation $\Delta\theta$ [$\Delta \phi$]).  In our fitting procedure, we set $p = 8$ for Type I and $p = 4$ for Type II and use the same standard deviation for the tube misalignment angles in fitting both Type I and II data.  The results of our fitting procedure are shown in Fig.~\ref{Experimental Fit}, where the best fits for Type I and Type II geometries are shown as dashed black and solid red lines, respectively.  Our best fit standard deviation is $\Delta \theta = \Delta \phi = 18.7^{\circ}$, which implies that the tube alignment angles on the sapphire substrate are $0^{\circ} \pm \ 9.35^{\circ}$.  For small $\Delta \theta$ (measured in radians) the nematic order parameter is $S = \exp(-2 \ (\Delta\theta)^2) = 0.81$.

Although a value of $S=0.81$ indicates a strongly aligned sample, it is not perfectly aligned ($S=1$). This is unlike previous work in the THz regime with these highly-aligned samples, where the nematic order parameter was calculated to be exactly $S=1$.\cite{Ren:2009p18188} It is clear that there is a wavelength dependence of the nematic order parameter, but we believe this can be explained qualitatively: any slight misalignment in the sample would go undetected in the THz regime, as the wavelength is much larger than our visible wavelengths that can detect the misalignments. This could explain the difference in calculated nematic order parameter between our CP measurements and previous THz measurements, but nonetheless, regardless of the calculated differences with wavelength, our calculated nematic order parameter is still quite large and we can confidently confirm the high degree of alignment of the sample with our CP measurements.

\section{Conclusion}

In summary, we investigated the polarization anisotropy of coherent phonon dynamics in highly-aligned single-walled carbon nanotubes and measured RBM coherent phonons as a function of polarization angle.  We saw a very nearly complete quenching of the RBM for both geometries and extended our results to determine the degree of alignment of the sample.  Comparing our results with theory, we performed simulations of polarization-dependent CP spectroscopy on a (38,0) zigzag nanotubes and also found a decrease in CP signal as optical polarization varies from parallel to perpendicular to the nanotube axis. Using those simulated results, we also theoretically determined a $\cos^8(\theta)$ dependence for Type I CP spectroscopy experiments and a $\cos^4(\phi)$ polarization dependence for Type II CP spectroscopy experiments. Including misalignment effects to our fitting, we finally determined the nematic order parameter of our sample to be 0.81.


\begin{acknowledgments}
This work was supported by the National Science Foundation under grant numbers DMR-0325474, OISE-0530220, and DMR-0706313,  the Robert A.~Welch foundation under grant number C-1509, and the Office of Naval Research (ONR) under contract number 00075094.  Y.-S. Lim and K.-J. Yee are supported by a Korea Science and Engineering Foundation (KOSEF) grant funded by the Korean Government (Most) (R01-2007-000-20651-0). We acknowledge useful discussions with Andrew Rinzler at the University of Florida.
\end{acknowledgments}

\bibliography{paper}


\end{document}